# Is there a metamaterial route to high temperature superconductivity?

Igor I. Smolyaninov [1)] and Vera N. Smolyaninova [2)]

[1]Department of Electrical and Computer Engineering, University of Maryland, College Park, MD 20742, USA

[2]Department of Physics Astronomy and Geosciences, Towson University,

8000 York Rd., Towson, MD 21252, USA

**Superconducting properties of a material, such as electron-electron interactions and the critical temperature of superconducting transition can be expressed via the effective dielectric response function $\varepsilon_{eff}(q,\omega)$ of the material. Such a description is valid on the spatial scales below the superconducting coherence length (the size of the Cooper pair), which equals ~100 nm in a typical BCS superconductor. Searching for natural materials exhibiting larger electron-electron interactions constitutes a traditional approach to high temperature superconductivity research. Here we point out that recently developed field of electromagnetic metamaterials deals with somewhat related task of dielectric response engineering on sub-100 nm scale. We argue that the metamaterial approach to dielectric response engineering may considerably increase the critical temperature of a composite superconductor-dielectric metamaterial.**

Electromagnetic properties are known to play a very important role in the pairing mechanism and charge dynamics of high $T_c$ superconductors [1]. Moreover, shortly after the original work by Bardeen, Cooper and Schrieffer (BCS) [2], Kirzhnits *et al.* formulated a complementary description of superconductivity in terms of the dielectric response function of the superconductor [3]. The latter work was motivated by a simple argument that phonon-mitigated electron-electron interaction in superconductors may be expressed in the form of effective Coulomb potential

$$V(\vec{q},\omega) = \frac{4\pi e^2}{q^2 \varepsilon_{eff}(\vec{q},\omega)}, \quad (1)$$

where $V=4\pi e^2/q^2$ is the usual Fourier-transformed Coulomb potential in vacuum, and $\varepsilon_{eff}(q,\omega)$ is the linear dielectric response function of the superconductor treated as an effective medium. Based on this approach, Kirzhnits et al. derived simple expressions for the superconducting gap $\Delta$, critical temperature $T_c$, and other important parameters of the superconductor. While thermodynamic stability condition implies [4] that $\varepsilon_{eff}(q,0) > 0$, the dielectric response function at higher frequencies and spatial momenta is large and negative, which accounts for the weak net attraction and pairing of electrons in the superconducting condensate. In their paper Kirzhnits *et al.* noted that this effective medium consideration assumes "homogeneous system" so that "the influence of the lattice periodicity is taken into account only to the extent that it may be included into $\varepsilon_{eff}(q,\omega)$".

In the forty years which had passed since this very important remark, we have learned that the "homogeneous system" approximation may remain valid even if the basic structural elements of the material are not simple atoms or molecules. Now we know that artificial "metamaterials" may be created from much bigger building blocks, and the electromagnetic properties of these fundamental building blocks ("meta-atoms") may be engineered at will [5]. Since the superconducting coherence length (the size of the Cooper pair) is $\xi$~100 nm in a typical BCS superconductor, we have an opportunity to engineer the fundamental metamaterial building blocks in such a way that the effective electron-electron interaction (1) will be maximized, while homogeneous treatment of $\varepsilon_{eff}(q,\omega)$ will remain valid. In order to do this, the metamaterial unit size must fall within a rather large window between ~0.3 nm (given by the atomic scale) and $\xi$~100 nm scale of a typical Cooper pair. However, this task is much more challenging than typical applications of superconducting metamaterials suggested so far [6], which only deal with metamaterial engineering on the scales which are much smaller than the microwave or RF wavelength. Our task requires development of superconducting metamaterials which are much more refined. In addition, the coherence length of the metamaterial superconductor must be determined in a self-consistent manner. The coherence length will decrease with increasing Tc of the metamaterial superconductor, since the approach of Kirzhnits et al. gives rise to the same BCS-like relationship between the superconducting gap $\Delta$ and the coherence length $\xi$ [3]:

$$\Delta\left(\frac{\xi}{V_F}\right) \sim \hbar, \qquad (2)$$

where $V_F$ is the Fermi velocity. Therefore, metamaterial structural parameter (such as the inter-layer distance, etc.) which must remain smaller than the coherence length, will define the limit of critical temperature increase.

Let us demonstrate that tuning electron-electron interaction is indeed possible in a metamaterial scenario. It is obvious from eq.(1) that the most natural way to increase attractive electron-electron interaction is to reduce the absolute value of $\varepsilon_{eff}(q,\omega)$ while keeping $\varepsilon_{eff}(q,\omega)$ negative within a substantial portion of the relevant four-momentum spectrum ($|\vec{q}| \leq 2k_F$, $\omega \leq$ BCS cutoff around the Debye energy). Potentially, this may be done using the epsilon-near-zero (ENZ) metamaterial approach [7], which is based on intermixing metal and dielectric components in the right proportions. A negative $\varepsilon \approx 0$ ENZ metamaterial would maximize attractive electron-electron interaction given by eq.(1). Let us consider a random mixture of superconducting "matrix" and dielectric "inclusions" described in the frequency range of interest by the dielectric constants $\varepsilon_m$ and $\varepsilon_i$, respectively. In the Maxwell-Garnett approximation the effective dielectric constant $\varepsilon_{eff}$ of the metamaterial may be obtained as

$$\left(\frac{\varepsilon_{eff}-\varepsilon_m}{\varepsilon_{eff}+2\varepsilon_m}\right)=\delta_i\left(\frac{\varepsilon_i-\varepsilon_m}{\varepsilon_i+2\varepsilon_m}\right), \quad (3)$$

where $\delta_i$ is the volume fraction of the inclusions (considered to be small) [8]. The explicit expression for $\varepsilon_{eff}$ may be written as

$$\varepsilon_{eff}=\varepsilon_m\frac{(2\varepsilon_m+\varepsilon_i)-2\delta_i(\varepsilon_m-\varepsilon_i)}{(2\varepsilon_m+\varepsilon_i)+\delta_i(\varepsilon_m-\varepsilon_i)} \quad (4)$$

The ENZ conditions ($\varepsilon_{eff} \approx 0$) are obtained around

$$\delta_i=\frac{2\varepsilon_m+\varepsilon_i}{2(\varepsilon_m-\varepsilon_i)}, \quad (5)$$

which means that $\varepsilon_m$ and $\varepsilon_i$ must have opposite signs, and $\varepsilon_i \approx -2\varepsilon_m$ so that $\delta_i$ will be small. This simple consideration indicates that attractive electron-electron interaction in a superconducting metamaterial may indeed be increased by using the correct amount of dielectric. However, $\varepsilon_i$ of the dielectric needs to be very large, since $\varepsilon_m$ of the metal component typically given by the Drude model in the far infrared and THz ranges

$$\varepsilon_m=\varepsilon_{m\infty}-\frac{\omega_p^2}{\omega^2}\approx-\frac{\omega_p^2}{\omega^2} \quad (6)$$

is large and negative (where $\varepsilon_{m\infty}$ is the dielectric permittivity of metal above the plasma edge and $\omega_p$ is its plasma frequency). Ferroelectric materials having large positive $\varepsilon_i$ in the same frequency ranges may be a very good choice of such dielectrics. On the other hand, Maxwell-Garnett based analysis of eqs. (1) and (4) indicates that even far from ENZ conditions a superconductor-dielectric metamaterial must have larger Δ and higher Tc compared to the original undiluted superconducting host. Indeed, even in the limit $\varepsilon_i << -\varepsilon_m$ and small $\delta_i$ Maxwell-Garnett approximation (eq.(4)) results in decrease of $\varepsilon_{eff}$

$$\varepsilon_{eff} = \varepsilon_m \frac{(2\varepsilon_m + \varepsilon_i) - 2\delta_i(\varepsilon_m - \varepsilon_i)}{(2\varepsilon_m + \varepsilon_i) + \delta_i(\varepsilon_m - \varepsilon_i)} \approx \varepsilon_m \frac{2\varepsilon_m - 2\delta_i\varepsilon_m}{2\varepsilon_m + \delta_i\varepsilon_m} \approx \varepsilon_m \left(1 - \frac{3}{2}\delta_i\right) \qquad (7)$$

producing an increase of

$$V(\vec{q},\omega) = \frac{4\pi e^2}{q^2 \varepsilon_{eff}(\vec{q},\omega)} \approx \frac{4\pi e^2}{q^2 \varepsilon_m(\vec{q},\omega)}(1 + \frac{3}{2}\delta_i) \qquad (8)$$

which should be proportional to the increase in $\Delta$ and Tc, and which in this limit does not depend on the particular choice of dielectric $\varepsilon_i$. According to eq.(2), compared to the parent pure superconductor, the metamaterial coherence length $\xi_{MM}$ will decrease as

$$\xi_{MM} = \frac{\xi_0}{\left(1 + \frac{3}{2}\delta_i\right)} \qquad (9)$$

where $\xi_0$ is coherence length of the parent superconductor. Therefore, the metamaterial design should make sure that the typical metamaterial structural parameter remains much below $\xi_{MM}$. This result constitutes a prof of principle of the metamaterial approach and can be confronted with experiments. While ENZ conditions must lead to considerably larger $Tc$ increases, detailed calculations of $Tc$ increase are much more difficult, since they require detailed knowledge of $\varepsilon_m(q,\omega)$ and $\varepsilon_i(q,\omega)$ of the metamaterial components. On the other hand, evaluation of the maximum critical temperature $T_c^{max}$ of the superconducting transition from the point of view of electromagnetic approach performed by Kirzhnits et al. in ref. [3] produced a very optimistic $T_c^{max}$~300K estimate at $E_F$~10eV. According to eq.(2), the corresponding $\xi_{MM}$ ~48 nm leaves substantial room for metamaterial engineering. As far as the London penetration length

$$\lambda_L = \left(\frac{m}{\mu_0 n e^2}\right)^{1/2} \qquad (10)$$

of such a composite superconductor is concerned, where $m$ is the effective mass, and $n$ is the carrier concentration, it is obvious that compared to undiluted superconductor the carrier concentration is smaller, leading to increased penetration length of the composite superconductor. Decrease of $\xi$ accompanied by increase of $\lambda_L$ will lead to composite metamaterial exhibiting type-II superconductivity.

Another interesting possibility is to use a hyperbolic metamaterial geometry shown in Fig.1. As has been noted in ref.[9], typical high $T_c$ superconductors (such as BSCCO) do exhibit hyperbolic metamaterial behavior in a substantial portion of far infrared and THz frequency ranges. Hyperbolic metamaterials are typically composed of multilayer metal-dielectric or metal wire array structures. However, a few natural materials, such as sapphire and bismuth [10] also exhibit hyperbolic behaviour in a limited frequency range. The diagonal components of dielectric permittivity tensor $\varepsilon_{xx}=\varepsilon_{yy}=\varepsilon_1$ and $\varepsilon_{zz}=\varepsilon_2$ of these non-magnetic uniaxial metamaterials have opposite signs, resulting in such unusual electromagnetic properties as absence of diffraction limit [11], and diverging photonic density of states [12]. Let us demonstrate that hyperbolic metamaterial geometry offers another natural way to increase attractive electron-electron interaction in a layered dielectric-superconductor metamaterial. However, we should emphasize that our theoretical consideration presented below is not intended to be an alternative theory of superconductivity in high Tc cuprates.

Since hyperbolic metamaterials exhibit considerable dispersion, let us work in the frequency domain and write macroscopic Maxwell equations in the presence of "external" electron density $\rho_\omega$ and current $J_\omega$ as

$$\frac{\omega^2}{c^2}\vec{D}_\omega = \vec{\nabla}\times\vec{\nabla}\times\vec{E}_\omega - \frac{4\pi i\omega}{c^2}\vec{J}_\omega, \quad \vec{\nabla}\cdot\vec{D}_\omega = \rho_\omega, \quad \text{and} \quad \vec{D}_\omega = \vec{\varepsilon}_\omega \vec{E}_\omega \quad (11)$$

where the frequency $\omega$ is assumed to fall within the hyperbolic frequency band of the metamaterial. Let us solve eq.(11) for the z-component of electric field. After straightforward transformations we obtain

$$\frac{\omega^2}{c^2}E_z = \frac{4\pi}{\varepsilon_1\varepsilon_2}\frac{\partial\rho}{\partial z} - \frac{4\pi i\omega}{c^2\varepsilon_2}J_z - \frac{\partial^2 E_z}{\varepsilon_1\partial z^2} - \frac{1}{\varepsilon_2}\left(\frac{\partial^2 E_z}{\partial x^2}+\frac{\partial^2 E_z}{\partial y^2}\right) \quad (12)$$

Since $E_z = \partial\phi/\partial z$, and the second term on the right side of eq.(9) may be neglected compared to the first one (since $v/c \ll 1$), we obtain

$$\frac{\omega^2}{c^2}\phi + \frac{\partial^2\phi}{\varepsilon_1\partial z^2}+\frac{1}{\varepsilon_2}\left(\frac{\partial^2\phi}{\partial x^2}+\frac{\partial^2\phi}{\partial y^2}\right) = \frac{4\pi}{\varepsilon_1\varepsilon_2}\rho \quad (13)$$

Taking into account that $V=-e\phi$, and neglecting the first term in eq(13) in the low frequency limit, we find that the effective Coulomb potential from eq.(1) assumes the form

$$V(\vec{q},\omega) = \frac{4\pi e^2}{q_z^2\varepsilon_2(\vec{q},\omega)+(q_x^2+q_y^2)\varepsilon_1(\vec{q},\omega)} \quad (14)$$

in a hyperbolic metamaterial. Since $\varepsilon_{xx} = \varepsilon_{yy} = \varepsilon_1$ and $\varepsilon_{zz} = \varepsilon_2$ have opposite signs, the effective Coulomb interaction of two electrons may become attractive and very strong in the hyperbolic frequency bands. The obvious condition for such a strong interaction to occur is

$$q_z^2 \varepsilon_2(\vec{q}, \omega) + \left(q_x^2 + q_y^2\right)\varepsilon_1(\vec{q}, \omega) \approx 0 \tag{15}$$

which indicates that the superconducting order parameter must be strongly anisotropic. This indeed appears to be the case in such hyperbolic high $T_c$ superconductors as BSCCO [1,9]. In order to be valid, the metamaterial "effective medium" description requires that the structural parameter of the metamaterial (in this particular case, the interlayer distance) must be much smaller than the superconducting coherence length. If the structural parameter approaches 1 nm scale, Josephson tunneling across the dielectric layers will become very prominent in such an anisotropic layered superconducting hyperbolic metamaterial.

Similar to the random superconductor-dielectric mixture considered above, the diagonal dielectric permittivity components of the layered superconductor-dielectric metamaterial may be calculated using Maxwell-Garnett approximation. These components can be calculated similar to [13] as follows:

$$\varepsilon_1 = \alpha \varepsilon_m + (1-\alpha)\varepsilon_d$$

$$\varepsilon_2 = \frac{\varepsilon_m \varepsilon_d}{(1-\alpha)\varepsilon_m + \alpha \varepsilon_d} \tag{16}$$

where $\alpha$ is the volume fraction of superconducting phase, and $\varepsilon_m < 0$ and $\varepsilon_d > 0$ are the dielectric permittivities of the superconductor and dielectric, respectively. If $\alpha$ is small, both $\varepsilon_1$ and $\varepsilon_2$ are positive. On the other hand, if $\alpha$ is close to 1, both $\varepsilon_1$ and $\varepsilon_2$ are negative. The hyperbolic conditions are obtained in the intermediate range of $\alpha$ if

$$((1-\alpha)\varepsilon_d + \alpha\varepsilon_m)\left(\frac{1-\alpha}{\varepsilon_d} + \frac{\alpha}{\varepsilon_m}\right) < 0 \tag{17}$$

After simple transformations the equations defining boundaries of the hyperbolic frequency band may be written as

$$\frac{\varepsilon_m(\omega)}{\varepsilon_d(\omega)} = -\frac{\alpha}{1-\alpha} \quad \text{and} \quad \frac{\varepsilon_m(\omega)}{\varepsilon_d(\omega)} = -\frac{1-\alpha}{\alpha} \tag{18}$$

Once again, $\varepsilon_d$ of the dielectric needs to be very large, since $\varepsilon_m$ of the superconducting component is negative and very large in the far infrared and THz ranges (see eq.(5)). This is consistent with the measured dielectric behavior of the parent BSCCO perovskite compound [1]. Moreover, if the high frequency behavior of $\varepsilon_d$ may be assumed to follow the Debye model [14]:

$$\text{Re}\,\varepsilon_d = \frac{\varepsilon_d(0)}{1+\omega^2\tau^2} \approx \frac{\varepsilon_d(0)}{\omega^2\tau^2}, \tag{19}$$

broadband hyperbolic behavior arise due to similar $\sim\omega^{-2}$ functional behavior of $\varepsilon_d$ and $\varepsilon_m$ in the THz range (compare eqs. (6) and (19)). Thus, BCS theory assumption of a constant attractive interaction from low frequencies (of the order of the gap energy) to the range of the BCS cutoff around the Debye energy would be approximately satisfied. However, this is not a strict requirement, and attractive interaction in metamaterial superconductors described by eqs.(1) or (13) may depend on frequency due to dispersive behavior of $\varepsilon_{eff}(q,\omega)$.

As follows from eq.(17), if the volume fraction of metallic phase $\alpha$ is kept constant, the hyperbolic behavior may occur only within the following range of plasma frequency $\omega_p^2 = ne^2/m^*$ of the metallic phase:

$$\frac{\omega_p^2 \tau^2}{\varepsilon_d(0)} \in \left[\frac{\alpha}{1-\alpha}; \frac{1-\alpha}{\alpha}\right], \qquad (20)$$

where $n$ is the free carrier (electron or hole) concentration, and $m^*$ is their effective mass. Otherwise, either $\varepsilon_1$ and $\varepsilon_2$ will be both positive if $\omega_p^2$ is too small, or both negative if $\omega_p^2$ is too large. Interestingly enough, the boundaries of superconducting and hyperbolic states in high Tc cuprates seem to overlap. Similar to hyperbolic behavior described by eq.(20), superconductivity occurs only within a certain narrow doping range $n_{min}<n<n_{max}$ of either electrons or holes [1]. Based on Fig.1 from ref.[1] the $n_{max}/n_{min}$ ratio equals 1.84 for the electron-doped $A_{2-x}Ce_xCuO_4$ compounds. On the other hand, based on the crystallographic lattice of BSCCO shown in Fig.1, $\alpha \approx 3/7$ in this material. Therefore, the ratio $n_{max}/n_{min}$ for the boundaries of hyperbolic phase given by eq.(20) is ~1.77. This close match appears to give us a hint at close relationship between the superconducting and hyperbolic behaviors in high Tc cuprates. Indeed, in BSCCO the anisotropy of DC conductivity may reach $10^4$ for the ratio of in plane to out of plane conductivity in high quality single crystal samples. Polarization-dependent AC reflectance spectra measured in the THz and far-infrared frequency ranges [1] also indicate extreme anisotropy. In the normal state of high Tc superconductors the in-plane AC conductivity exhibits Drude-like behavior with a plasma edge close to 10000cm$^{-1}$, while AC conductivity perpendicular to the copper oxide planes is nearly insulating. Extreme anisotropy is also observed in the superconducting state. The typical values of measured in-plane and out of plane condensate plasma frequencies in high Tc superconductors are $\omega_{s,ab}$=4000-10000cm$^{-1}$, and $\omega_{s,c}$=1-1000cm$^{-1}$, respectively [1]. The measured anisotropy is the strongest in the BSCCO superconductors. These experimental measurements strongly support the qualitative picture of BSCCO structure as a layered hyperbolic metamaterial (Fig.1(b)) in which the copper oxide layers may be represented as metallic layers, while the SrO and BiO layers may be represented as the layers of dielectric. Based on these measured material parameters, we may calculate the diagonal components of BSCCO dielectric tensor.

Our calculations will be performed for $T<T_c$ and centred around the spectral range $\omega_{s,c}<\omega<\omega_{s,ab}$, so that AC conductivity perpendicular to the copper oxide layers may indeed be considered dielectric. For the in-plane permittivity we assume Drude-like behavior supported by AC measurements (see Fig.5 from [1]):

$$\varepsilon_m = \varepsilon_{m\infty} - \frac{\omega_{s,ab}^2}{\omega^2} \quad , \tag{21}$$

while the out of plane permittivity will be approximates as

$$\varepsilon_d = \varepsilon_{d\infty} - \frac{\omega_{s,c}^2}{\omega^2} \quad , \tag{22}$$

where $\varepsilon_{m\infty} \sim 4$ is the dielectric permittivity of copper oxide layers above the plasma edge, and $\varepsilon_{d\infty}$ is the dielectric permittivity of the undoped insulating parent copper oxide compound. The parent perovskite compounds typically exhibit rather large values of $\varepsilon_{d\infty}$, which may be estimated from their out of plane reflectivity. Based on Fig.5 from [1] $\varepsilon_{d\infty} \sim 25$ may be assumed. The calculated diagonal components of the permittivity tensor for a high $T_c$ superconductor having $\omega_{s,ab}$=10000cm$^{-1}$ and $\omega_{s,c}$=1000cm$^{-1}$ are presented in Fig.2. Based on the crystallographic unit cell of BSCCO $\alpha$=0.36 has been assumed in these calculations. The hyperbolic behavior appears in the very broad 200cm$^{-1}$<$\omega$<1200cm$^{-1}$ and 2400cm$^{-1}$<$\omega$<4800cm$^{-1}$ spectral ranges. Within these spectral bands $\varepsilon_1$ and $\varepsilon_2$ have opposite signs. The appearance of hyperbolic bands is quite generic, independent of a particular choice of $\omega_{s,ab}$ and $\omega_{s,c}$, as long as strong anisotropy of $\omega_s$ is maintained. While examples of such natural hyperbolic high $T_c$ superconductors appear to fit well into the metamaterial scheme described above, it would be interesting to try and follow the metamaterial recipe in making novel "designer" superconductors.

Since metamaterial dimensions required for engineering of electron-electron interaction approach nanometer scale, another potentially important issue is applicability of nanoscale metal and dielectric layers description using their macroscopic dielectric constants. This issue is well known and extensively studied in nanophotonics and electromagnetic metamaterials. The electromagnetic response of thin metal layers is indeed known to exhibit weak oscillatory dependence on layer thickness due to quantum mechanical effects, such as formation of electron standing waves inside the thin layer [15]. While this effect indeed affects effective dielectric constant of a metal layer, for all practical purposes this is a weak effect. On the other hand, the dielectric constant of dielectric materials does not depend on layer dimensions until the atomic scale is reached. This fact has been verified in experiments on surface plasmon resonance [16].

In conclusion, we pointed out that recent developments in the field of electromagnetic metamaterials, such as development of epsilon near zero (ENZ) and hyperbolic metamaterials may be used to engineer dielectric response of composite superconducting metamaterials on sub-100 nm scale. We argue that the metamaterial approach to dielectric response engineering may considerably increase the critical temperature of such composite superconductor-dielectric metamaterials.

**References**

[1] D.N. Basov and T. Timusk, "Electrodynamics of high-$T_c$ superconductors", *Rev. Mod. Phys.* **77**, 721 (2005).

[2] J. Bardeen, L.N. Cooper, and J.R. Schrieffer, "Theory of superconductivity", Phys. Rev. 108, 1175 (1957).

[3] D.A. Kirzhnits, E.G. Maksimov, and D.I. Khomskii, "The description of superconductivity in terms of dielectric response function", *J. Low Temp. Phys.* **10**, 79 (1973).

[4] L.D. Landau and E.M. Lifshitz, *Macroscopic Electrodynamics* (Elsevier, Oxford, 2000).

[5] J.B. Pendry, D. Schurig, and D.R. Smith, "Controlling electromagnetic fields", *Science* **312**, 1780 (2006).

[6] S.M. Anlage, "The physics and applications of superconducting metamaterials", *J. Opt.* **13**, 024001 (2011).

[7] N. Engheta, "Pursuing near-zero response", *Science* **340**, 286 (2013).

[8] T.C. Choy, *Effective Medium Theory* (Clarendon Press, Oxford, 1999).

[9] I. I. Smolyaninov, "Quantum nucleation of effective Minkowski spacetime in hyperbolic metamaterials based on high Tc superconductors", arXiv:1306.1975

[10] L. V. Alekseyev, V. A. Podolskiy, and E. E. Narimanov, *Advances in OptoElectronics* **2012**, 267564 (2012).

[11] Z. Jakob, L.V. Alekseyev, E. Narimanov, "Optical hyperlens: far-field imaging beyond the diffraction limit", *Optics Express* **14,** 8247 (2006).

[12] I. I. Smolyaninov and E. E. Narimanov, "Metric signature transitions in optical metamaterials", *Phys. Rev. Letters* **105**, 067402 (2010).

[13] R. Wangberg, J. Elser, E.E. Narimanov, and V.A. Podolskiy, "Nonmagnetic nanocomposites for optical and infrared negative-refractive-index media", *J. Opt. Soc. Am. B* **23**, 498-505 (2006).

[14] C. Kittel, *Introduction to Solid State Physics* (Wiley, New York, 2004).

[15] M. Cini and P. Ascarelli, "Quantum size effects in metal particles and thin films by an extended RPA", *J. Phys. F: Metal Phys*. **4**, 1998-2008 (1974).

[16] A.V. Zayats, I.I. Smolyaninov, and A. Maradudin, "Nano-optics of surface plasmon-polaritons", *Physics Reports* **408**, 131-314 (2005).

**Figure Captions**

**Figure 1.** Comparison of the crystallographic unit cell of a BSCCO high $T_c$ superconductor (a) and geometry of a layered hyperbolic metamaterial (b).

**Figure 2**. Diagonal components of the permittivity tensor of a high Tc BSCCO superconductor calculated as a function of frequency assuming $\omega_{s,ab}$=10000cm$^{-1}$, $\omega_{s,c}$=1000cm$^{-1}$, $\varepsilon_{m\infty} = 4$ and $\varepsilon_{d\infty} = 25$. The hyperbolic bands appear at 200cm$^{-1}$<$\omega$<1200cm$^{-1}$ and 2400cm$^{-1}$<$\omega$<4800cm$^{-1}$.

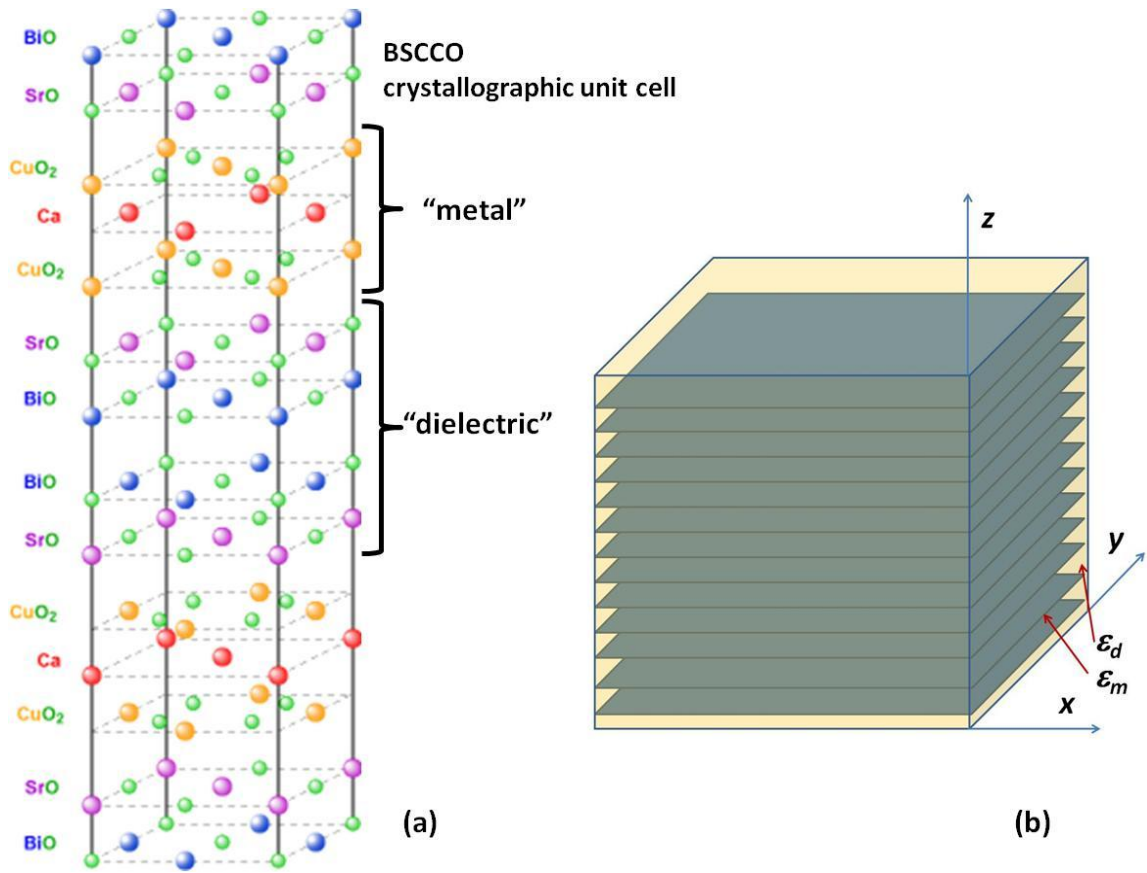

Fig.1

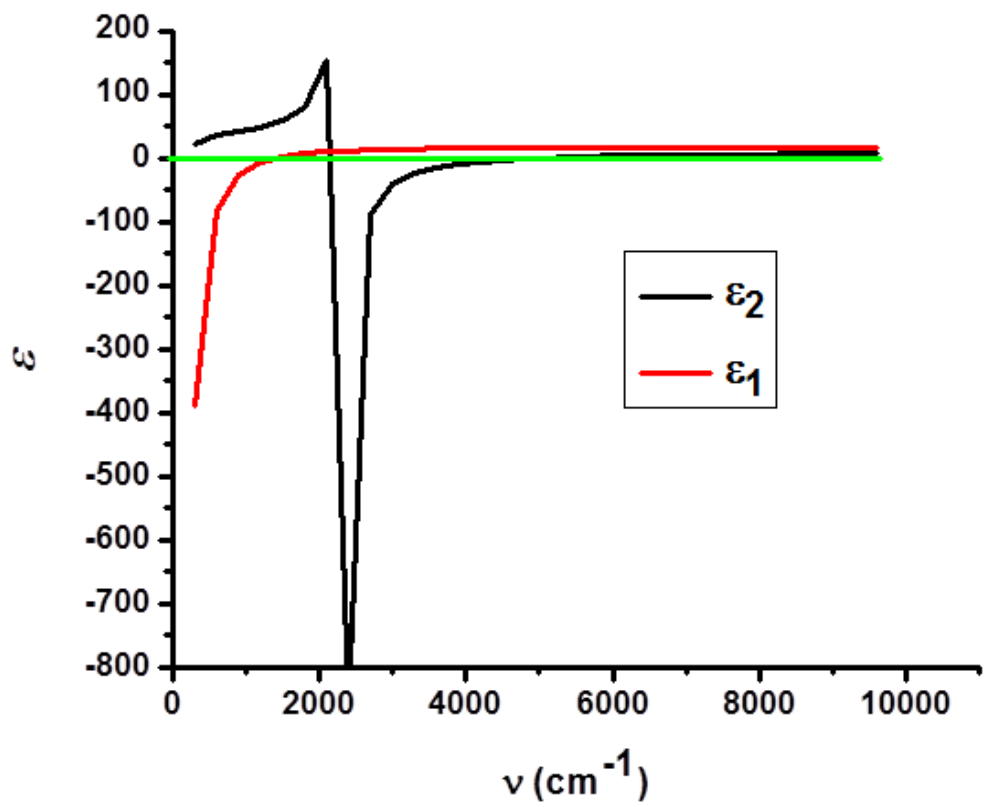

Fig. 2